\documentclass[sigconf,screen,nonacm]{acmart}

\expandafter\def\expandafter\UrlBreaks\expandafter{\UrlBreaks\do\/\do\*\do\-\do\~\do\'\do\"\do\-}
\usepackage{color}

\usepackage[shortcuts]{extdash}
\usepackage{color,soul}
\usepackage{multirow}
\usepackage{graphicx}
\usepackage{listings}
\usepackage[normalem]{ulem}

\graphicspath{ {./media/} }

\newcommand{\revise}[1]{{#1}}

\usepackage{enumitem}

\begin{document}

\title{Running Cloud-native Workloads on HPC with High-Performance Kubernetes}

\author{Antony Chazapis}
\affiliation{%
    \institution{FORTH}
    \department{Institute of Computer Science}
    \city{Heraklion}
    \country{Greece}}
\email{chazapis@ics.forth.gr}

\author{Evangelos Maliaroudakis}
\authornote{Also with University of Crete, Computer Science Department.}
\affiliation{%
    \institution{FORTH}
    \department{Institute of Computer Science}
    \city{Heraklion}
    \country{Greece}}
\email{malvag@ics.forth.gr}

\author{Fotis Nikolaidis}
\affiliation{%
    \institution{FORTH}
    \department{Institute of Computer Science}
    \city{Heraklion}
    \country{Greece}}
\email{fnikol@ics.forth.gr}

\author{Manolis Marazakis}
\affiliation{%
    \institution{FORTH}
    \department{Institute of Computer Science}
    \city{Heraklion}
    \country{Greece}}
\email{maraz@ics.forth.gr}

\author{Angelos Bilas}
\authornotemark[1]
\affiliation{%
    \institution{FORTH}
    \department{Institute of Computer Science}
    \city{Heraklion}
    \country{Greece}}
\email{bilas@ics.forth.gr}

\begin{abstract}
The escalating complexity of applications and services encourages a shift towards higher-level data processing pipelines that integrate both Cloud-native and HPC steps into the same workflow. Cloud providers and HPC centers typically provide both execution platforms on separate resources. In this paper we explore a more practical design that enables running unmodified Cloud-native workloads directly on the main HPC cluster, avoiding resource partitioning and retaining the HPC center's existing job management and accounting policies.
\end{abstract}

\settopmatter{printacmref=false, printccs=false, printfolios=true}

\maketitle
\renewcommand{\shortauthors}{A. Chazapis, E. Maliaroudakis, F. Nikolaidis, M. Marazakis, and A. Bilas}

\section{Introduction}

Cloud-HPC convergence
for Big Data processing pipelines that combine Cloud-native with HPC steps
is most often realized with interfacing mechanisms for submitting HPC jobs from the Cloud side or vice versa. However, bridging separate Cloud and HPC environments has proven to be challenging, as it involves synchronizing data between sites and coping with authorization restrictions. Having two separate setups also elevates the associated hardware and maintenance costs.

In this paper, we explore an HPC-centric solution that accommodates both Cloud and HPC software stacks on the same physical resources.
We focus our work on Kubernetes, currently the most prominent distributed container orchestrator for supporting the ``Cloud-native'' ecosystem.
We present \textit{High-Performance Kubernetes} (HPK), an open-source integration of unmodified Kubernetes components and custom modules that runs 
as a user-level service, requiring minimal support from the underlying HPC system software \revise{{\cite{hpk-paper, hpk}}}.
With HPK, users run their own private ``mini Clouds'' as in-cluster applications that delegate scheduling decisions to Slurm and execution to the Singularity/Apptainer container runtime, to comply with organization policies and established resource accounting mechanisms.
We evaluate HPK by running unmodified Cloud-native data analysis and machine learning applications in an HPC cluster, showcasing how it enables users to tap on the vast collection of available data processing frameworks, both as stand-alone solutions and in hybrid computation scenarios alongside HPC 
codes.


\section{Related work}
\label{sec:related}

We classify work related to Cloud-HPC convergence in two main categories: Systems that maintain the \emph{separation} of Cloud and HPC, and systems that \emph{embed} one resource management framework into the other.
In the former case there are two separate resource managers, while in the latter there is a single authority that controls hardware allocations, shared by both Cloud and HPC deployments; the embedded framework delegates resource management decisions to the overall cluster manager.
Works that assume separate clusters can further be divided into \emph{bridging} solutions that operate within the context of the Cloud or HPC runtime, allowing the transparent submission of remote jobs, or \emph{third-party} systems that operate in their own context and are able to administer tasks in both remote Cloud and HPC installations.

Many bridging solutions are available for Kubernetes, enabling Cloud users to integrate the execution of remote HPC tasks into their workflows. The \emph{hpc-connector} \cite{10.1007/978-3-030-59851-8_20} uses a custom container that when executed as part of a Kubernetes Job, will forward work to the HPC cluster, track its execution, and collect any results.
Other tools use Kubernetes Custom Resource Definitions (CRDs) that determine how to describe jobs targeted for the HPC cluster.
\emph{Torque-Operator} \cite{9284299} interfaces a Kubernetes installation to a Torque-based HPC cluster, while the \emph{Bridge Operator} \cite{lublinsky2022kubernetes} offers a wider compatibility of remote job execution facilities.
To avoid CRDs, KNoC (Kubernetes Node on HPC Cluster)~\cite{knoc} implements a virtual node for Kubernetes that transparently manages the container lifecycle on a remote HPC cluster using Slurm and Singularity.
This technique effectively allows users to employ existing Cloud-native tools, such as Argo Workflows to express complex data-processing pipelines for both Cloud and HPC without explicit remote execution steps.

Bridging solutions are especially useful when Cloud and HPC resources are colocated. HPC centers increasingly support on-demand provisioning of Cloud resources---even as partitions of the main HPC machine. However, when the two are remotely situated, bridging suffers from the overhead of maintaining data copies.
The user must prepare and send inputs to the remote HPC cluster before issuing any tasks, and then place back outputs in the Kubernetes context.
Data synchronization in hybrid workflows is addressed by StreamFlow~\cite{9177340}, a third-party system which extends the workflow language with declarative descriptions of execution sites (either Cloud or HPC) and their relationship to workflow nodes. The runtime automatically infers data dependencies, so to copy required data where needed before running each step.

Embedded convergence solutions avoid data copies and the requirement to manage and maintain two separate setups, as both Cloud and HPC share a common hardware platform.
The MPI Operator \cite{mpi-operator} defines the \texttt{MPIJob} Kubernetes CRD, which is realized by running a group of ephemeral container instances that include the MPI process.
\emph{Virtual clusters} \cite{virtual-clusters} take this notion a step further, as the execution containers embed a Slurm deployment for compatibility with existing scripts.
Additionally, the Slurm controller is extended with a custom protocol that requests resources from the Kubernetes scheduler, effectively maintaining Kubernetes's role of managing the whole infrastructure. A custom Kubernetes scheduler allows applying different container placement policies for ``HPC'' and ``data center'' services (Kubernetes deployments that run in other containers). \revise{SUNK {\cite{sunk}} (SlUrm oN Kubernetes) is based on a similar idea, although using a cluster-wide Slurm deployment that is integrated into Kubernetes as a specialized scheduler. External Slurm ``login nodes'' provide users with the typical interface of running Slurm scripts, while actual jobs are submitted to the Kubernetes cluster nodes. A ``syncer'' plugin monitors resource utilization to synchronize state information between Kubernetes and Slurm.}

Embedded configurations have also been classified as \cite{slurmAndOrVsKubernetes}: \emph{Over}, where Slurm is in control of the cluster, creating Kubernetes environments ephemerally within batch jobs, \emph{adjacent}, when both Slurm and Kubernetes are installed on the same physical nodes but share a common scheduler (i.e., Kubernetes uses Slurm to place jobs), and \emph{under}, when Slurm-enabled pods are deployed in Kubernetes (like virtual clusters). HPK falls within the \emph{over} class, however it does not create a full Kubernetes environment as a batch job (or as a dynamic partition, as may be possible using Flux child instances \cite{flux}), but rather transforms each Kubernetes deployment to an individual Slurm script, allowing for better scheduling flexibility and finer-grain resource sharing.

We are not aware of any other system that embeds Kubernetes in HPC in such a way. Usernetes \cite{usernetes} is a step in this direction, providing a Kubernetes distribution that can run without root privileges.
We did consider extending Usernetes to implement HPK, but quickly realized that the necessity of interfacing with Slurm and Singularity/Apptainer at multiple levels, would require reevaluating the internal structure of Kubernetes leading to reimplementing several subsystems. Interestingly, Usernetes solves the problem of managing the system's routing tables by utilizing a user-level networking stack, although this imposes several requirements to the environment, including availability of specific kernel modules.

\section{Design \& implementation}
\label{sec:design}

\begin{figure*}[thb]
    \centering
    \includegraphics[width=0.60\linewidth]{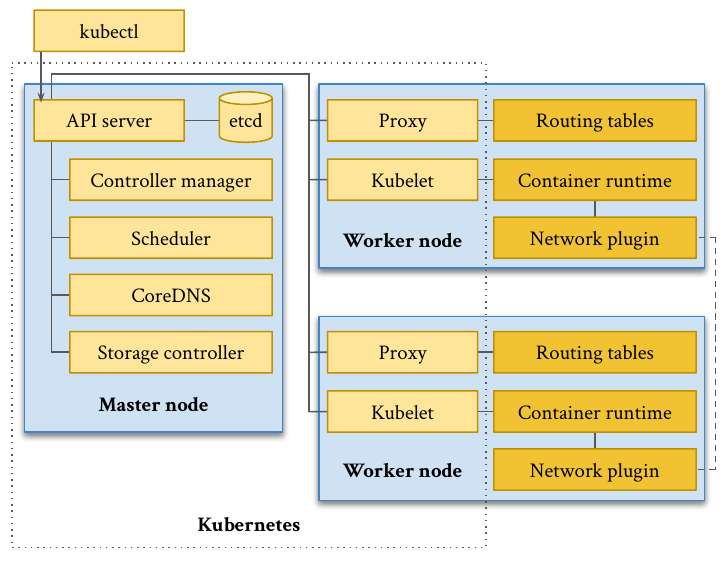}
    \caption{Components involved in a typical Kubernetes deployment on bare-metal.}
    \label{fig:kubernetes}
\end{figure*}

We aim to establish a framework that enables HPC users to execute Kubernetes workloads within a standard cluster environment.
From a design perspective, the requirements are:
\begin{itemize}
\item
{\it Compatibility}---\revise{To support the majority of Kubernetes applications and services, all native language abstractions, such as pods (one or more containers that are scheduled and scaled as a group), deployments, services, jobs, and volumes should be available and fully functional. Exceptions can only include constructs that directly relate to physical hardware resources (i.e., ``NodePort'' services that request a specific port number for exposing services at Kubernetes nodes). Microservices require that pods are individually addressable, supporting inter-container networking and internal service discovery.}
\item
{\it Compliance}---All resource management decisions should be delegated to the cluster manager operating the cluster (i.e., Slurm). Organization policies for resource allocation and accounting should be fully respected.
Running workloads should be visible at the level of the cluster manager.
Also, minimal configuration changes should be required to be done at the host level by HPC administrators. Reliance on special libraries or binaries that execute with ``elevated'' permissions should be avoided.
\item
{\it Usability}---Make it easy for users to deploy. All binaries should be packaged up with their dependencies into a container, with no host-specific requirements. All configuration should reside in the user's home directory.
\end{itemize}

Kubernetes is implemented as a set of communicating subsystems that collectively provide the functionality of distributed container orchestration. A typical deployment constitutes of the following (Figure \ref{fig:kubernetes}):
\begin{itemize}
\item
{\it API server}---The ``heart'' of Kubernetes. The main interface to the cluster and the synchronization point for all controllers.
\item
{\it etcd}---The key-value store holding all state. Always accessed through the API server.
\item
{\it Controller manager}---Watches for configuration changes or failures and performs all necessary actions to reach the desired state set by the user. The controller manager includes the controllers that implement the logic for the base Kubernetes abstractions. As all controllers, it communicates only with the API server.
\item
{\it Scheduler}---A controller that decides which node will be used to run new pods.
\item
{\it CoreDNS}---A controller and DNS server for implementing naming and discovery for pods and internal cluster services.
\item
{\it Kubelet}---An agent running on each worker node, implementing the pod lifecycle using a specific container runtime (i.e., \revise{containerd}).
\item
{\it Network plugin}---A service supporting the Container Network Interface (CNI) specification that assigns addresses to pods. The network plugin, which---depending on the Kubernetes version---is used by the kubelet or the container runtime directly, implements the Kubernetes network model. In addition to assigning unique, cluster-wide addresses, it makes sure that pods can communicate with each other across hosts. It may also realize traffic shaping policies or other network-level features.
\item
{\it Network proxy}---Creates local network routes for virtual IP addresses used by ``ClusterIP'' services. Runs on each worker node, alongside the kubelet.
\item
{\it Storage controller}---A controller that provisions storage of some type that can be attached to pods. Creates physical volumes to match requested persistent volume claims. While this controller is optional, we consider it a core component for a functioning system.
\end{itemize}

From these components, the \textit{API server}, \textit{etcd}, \textit{controller manager}, and \textit{CoreDNS} are self-contained services that can run as user processes. \revise{They are used unmodified in HPK, so from a user's perspective HPK provides the same frontend interface and API as any other Kubernetes deployment. Existing configurations using advanced language features, like isolated deployments in different namespaces and access control (RBAC) rules, can be directly applied without changes.}

\begin{figure}[tbh]
    \centering
    \includegraphics[width=0.65\linewidth]{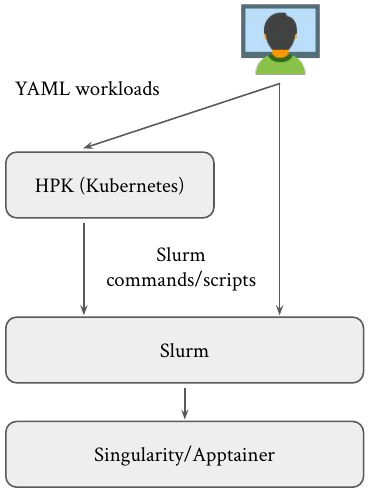}
    \caption{HPK translates Kubernetes workloads to Slurm and Singularity/Apptainer.}
    \label{fig:flow}
\end{figure}

To interface with the container runtime, HPK implements a custom \textit{kubelet}. \textit{hpk-kubelet} layers above both Slurm and Singularity/Apptainer for execution, translating the Kubernetes-level container lifecycle to Slurm scripts containing Singularity/Apptainer commands\revise{, while also synchronizing Kubernetes with the respective Slurm-level job states (i.e., enqueued jobs are marked as ``pending'' pods in Kubernetes, ``running'' when started, or ``failed'' if they produce errors).} With a single kubelet proxying requests to Slurm, the whole HPK integration can be visualized as a \textit{translation service}: Workloads enter in YAML format through the Kubernetes API endpoint and exit as Slurm scripts from hpk-kubelet (Figure \ref{fig:flow}).
\revise{We use generic Slurm directives in scripts that are not tied to a specific Slurm version.}
This design satisfies the \textit{compliance} requirement, as individual Kubernetes workloads transparently show up in Slurm queues. Moreover, Kubernetes-level resource requests (i.e., number of CPUs, amount of memory) are forwarded directly to Slurm, as are some HPK-specific pod annotations that can be used to further customize execution.
hpk-kubelet is implemented as a \textit{Virtual Kubelet Provider} \revise{{\cite{virtual-kubelet}}}. To respect pod network semantics (containers within the same pod share the same external IP address and can use \textit{localhost} to communicate with each other internally), it uses an embedded container topology: hpk-kubelet starts a ``parent'' container, which in turn runs each container of the pod. The pod IP address is assigned to the parent container; ``child'' containers run within the same network context without extra IP addresses.

In an HPK deployment, the Kubernetes cluster consists of just one worker node that represents the total amount of computing resources available \revise{to the user. Since cluster-level scheduling is to be performed by Slurm, HPK employs a custom, simplified \textit{pass-through scheduler} that makes no scheduling decisions, but always selects hpk-kubelet to run workloads.}

For pod and service networking, HPK avoids the \textit{network plugin} and \textit{network proxy} subsystems, as they perform actions at the system level as the root user.
Pod addresses should actually be assigned by the container runtime.
Singularity/Apptainer supports CNI plugins and can be easily set up to delegate network addressing to a cluster-wide service\revise{, which must be set up by system administrators at the node level} (i.e., Flannel).
This, in addition to allowing containers to run as \textit{fakeroot} for supporting common Docker images that use the root user, are the only changes HPK \revise{currently} requires from the HPC environment; both being configuration options of the container runtime.

To avoid the network proxy, HPK completely disables ``ClusterIP'' services, via a Kubernetes \textit{admission controller}---a hook that monitors API requests and may reject or mutate them before reaching the API server.
In Kubernetes, services can explicitly request to not use a virtual service IP (also called ``headless'' services). In such cases, service discovery continues to function, as CoreDNS maps the service name to the actual pod IPs instead of the virtual service address. Thus, microservice architectures are not affected by the lack of service-specific IPs. If load-balancing between pods is necessary, it can be implemented by using an additional service within a deployment.

\revise{For storage, HPK supports ``HostPath'' volumes---Kubernetes volumes that are bound to existing directories on the host. These can then be used by existing \textit{storage controllers}, such as OpenEBS, to provide container storage from different local or cluster-wide facilities through separate storage classes. As an example, users may deploy one OpenEBS storage class over node-local NVMe devices for temporary data, and another over their Lustre-backed home directory. Higher-level storage services, such as object stores, can then utilize these storage classes.}

\begin{figure*}[tbh]
    \centering
    \includegraphics[width=0.52\linewidth]{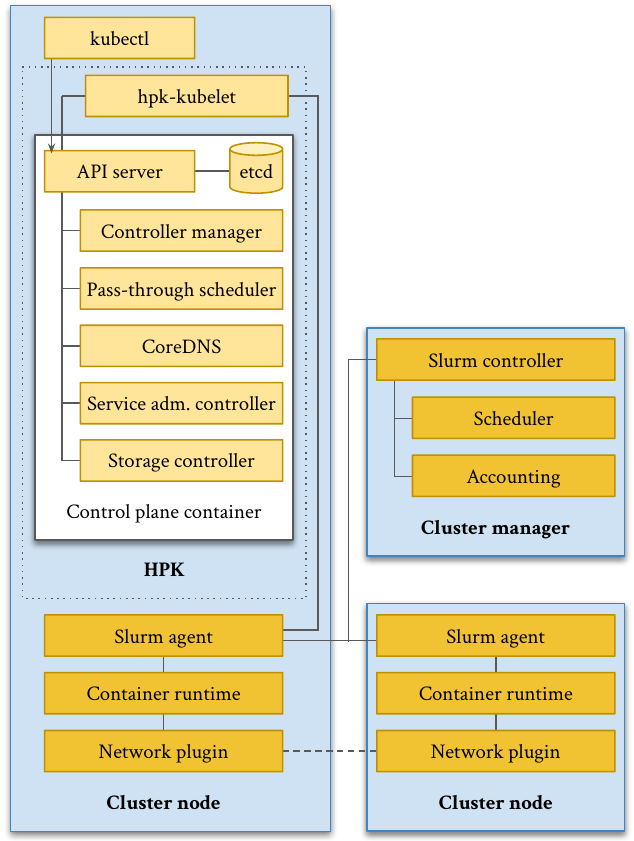}
    \caption{HPK architecture.}
    \label{fig:hpk}
\end{figure*}

The overall architecture of HPK is shown in Figure \ref{fig:hpk}. The \textit{control plane container} (in white background) includes all HPK components except the hpk-kubelet. It packs official releases of the API server, etcd, controller manager, and CoreDNS binaries \revise{(all corresponding to a specific Kubernetes version)}, along HPK's \revise{pass-through} scheduler and service admission controller. At runtime, it generates all necessary internal keys and certificates, bootstraps the Kubernetes control plane by initializing the executables in order, and produces the configuration file containing the endpoint and credentials needed to connect to the API server. Then, the hpk-kubelet uses this configuration to connect and announce its availability as a node.
\revise{HPK components run via Slurm with minimal resource requirements on any cluster node. Currently, the control plane container runs a single instance of each bundled service (i.e., API controller, etcd) and can be configured to resume its state in case it fails or expires due to exceeding the time limit imposed by the HPC site. We plan to address scalability and high-availability of the controle plane in future versions.}

\section{Evaluation}
\label{sec:evaluation}

To evaluate HPK's ability to efficiently deploy unaltered Cloud-native workloads on HPC, we use AWS's ParallelCluster, which provides a Cloud-based HPC-as-a-service environment managed by Slurm. We leverage the \texttt{OnNodeConfigured} action of ParallelCluster's configuration to execute a custom script upon node initialization in order to install additional software required by HPK. This includes Apptainer with the Flannel CNI plugin and Flannel, a tool to distribute private IPs to container instances and manage routes across nodes. Apptainer is set up to use Flannel for networking when running fakeroot containers (containers that internally use the root user), which is the default HPK configuration to allow seamless execution of Docker images.

For each experiment, we connect as a non-root user to the cluster's login node and run both HPK's control plane container, as well as hpk-kubelet.
By setting the \texttt{KUBECONFIG} environment variable to the configuration file produced, we can interface with HPK using common tools, such as \texttt{kubectl} and \texttt{helm}.

\subsection{Spark TPC-DS}
TPC-DS is an industry standard benchmark for measuring the performance of data processing systems. \revise{We run a sample integration provided by Amazon for EKS {\cite{eks-spark-benchmark}}, which uses Spark SQL orchestrated via the Spark Operator. Our goal is} to evaluate how HPK handles Spark-based workloads as they would typically be deployed in the Cloud. The operator streamlines the deployment and management of Apache Spark applications on Kubernetes by defining the \texttt{SparkApplication} CRD.
\revise{It} handles the entire lifecycle of execution, including submission, scaling, and cleanup, and provides logging and monitoring mechanisms for enhanced visibility into performance.

First, we deploy the Spark Operator and MinIO via Helm.
MinIO is an S3-compatible storage system, used to store the generated data. The benchmark requires a data generation phase before the actual submission of the workload. The benchmark YAMLs require that the S3 service is named \texttt{spark-k8s-data}, so this needs to be adjusted when deploying MinIO.

Once the supporting tools are ready, we submit the data generation and the benchmark \texttt{SparkApplications} in series, optionally adjusting the number executors and their resource requirements as shown in Listing \ref{datagen:tpcds} (the example YAMLs use 3 executors, each occupying 1 CPU core). The same \texttt{SparkApplication} YAMLs, without any changes, run in both a regular Cloud setting and HPK.

\begin{figure}[t]
\begin{lstlisting}[caption=Controlling the number of executors to be deployed during the Spark TPC-DS data generation step.,captionpos=b,label=datagen:tpcds,basicstyle=\scriptsize\ttfamily,numbers=left,xleftmargin=3em]
apiVersion: "sparkoperator.k8s.io/v1beta2"
kind: SparkApplication
metadata:
  name: tpcds-benchmark-data-generation-1g
spec:
  ...
  executor:
    instances: 3
    cores: 1
    memory: "8000m"
    memoryOverhead: 2g
\end{lstlisting}
\end{figure}

\subsection{Argo Workflows}
Argo Workflows, a Cloud-native workflow environment, provides a language (via a CRD) and runtime to model and execute applications as directed acyclic graphs (DAGs). In Argo, every node of the graph is a container. The Argo controller processes each workflow by submitting respective containers for execution, monitoring their status, and collecting their outputs; all presented via a user-friendly, interactive, web-based frontend. The frontend also allows organizing workflows using templates, as well as planning repeated execution with a crontab-like syntax.
HPK successfully runs the workflow examples that are included in the Argo repository, supporting all language features.

Additionally, similar to KNoC, HPK will pass-through specific pod annotations unmodified as additional Slurm flags. This also allows running a single container as a job in Kubernetes and scaling it up in the HPC environment via MPI parameters. The usage pattern is unusual for Cloud users, but may prove helpful in HPC for hybrid workflows, as it allows embedding MPI codes as steps and individually defining their scale. A simple example of an Argo workflow with an HPC step running the Embarrassingly Parallel NAS benchmark is shown in Listing \ref{code:nas}. We use the language's \texttt{\small{withItems}} construct to spawn 4 parallel steps, each running another instance of the executable with different parameters. An HPK Slurm annotation on the step template controls the number of tasks used for each instance. This showcases a method to run a parallel parameter sweep as part of a larger workflow. The ``items'' used may be explicitly set or be dynamically generated as the output of a previous step.

\begin{figure}[thb]
\begin{lstlisting}[caption={A simple Argo workflow executing multiple MPI steps in parallel, each with a different number of Slurm tasks.},captionpos=b,label=code:nas,basicstyle=\scriptsize\ttfamily,numbers=left,xleftmargin=3em]
kind: Workflow
metadata:
  ...
spec:
  entrypoint: npb-with-mpi
  templates:
  - name: npb-with-mpi
    dag:
      tasks:
      - name: A
        template: npb
        arguments:
          parameters:
          - {name: cpus, value: "{{item}}"}
        withItems:
        - 2
        - 4
        - 8
        - 16
  - name: npb
    metadata:
      annotations:
        slurm-job.hpk.io/flags: >-
          "--ntasks={{inputs.parameters.cpus}}"
        slurm-job.hpk.io/mpi-flags: "..."
    inputs:
      parameters:
      - name: cpus
    container:
      image: mpi-npb:latest
      command: ["ep.A.{{inputs.parameters.cpus}}"]
\end{lstlisting}
\end{figure}

\subsection{Distributed ML Training}
Argo Workflows is also extensively used for orchestrating multi-stage ML training pipelines (stand-alone or as part of the Kubeflow toolkit). We evaluate a workflow that trains several different models to classify images from the Fashion MNIST dataset, before selecting the one with the best accuracy to support an inference service \cite{distributed-ml-system}. The workflow leverages the Training Operator from Kubeflow to coordinate the execution of distributed training with TensorFlow. Instead of simple container image steps, it uses \texttt{TFJob} CRDs; the operator then spawns the requested number of pods with the appropriate roles (i.e., parameter servers, workers) and handles their lifecycle.

In addition to Argo Workflows, we install the Training Operator from the Kubeflow project. We then run a simplified version of the workflow that includes only the first two steps: the initial data ingestion and the distributed training of the models.
The preparatory phase also ensures that datasets are suitably formatted and environments are configured for the impending  training.
For training, the workflow utilizes the \texttt{MultiWorkerMirroredStrategy} with Keras, which implements synchronous distributed training across multiple workers.
Both steps run without issues in HPK, verifying its compatibility with higher-level Cloud-native tools. The level of automation and abstraction provided by the combination of Argo Workflows with the Training Operator becomes particularly valuable when constructing complex computational pipelines with multiple parallel steps, also involving decision making in the process.

\section{Conclusion}
\label{sec:conclusion}

Kubernetes has become the industry standard runtime in the Cloud, providing the necessary abstractions to
embrace the breadth and heterogeneity of available resources. Compatible Cloud-native tools are constantly
evolving, covering a wide spectrum of applications, including database and queuing systems, interactive code
execution frontends, workflow management utilities, as well as development frameworks that automatically
optimize and scale operations.
The ability to deploy this software directly on an HPC cluster via HPK opens up new possibilities for both Cloud and HPC users.
HPK can be used to attract Cloud users to large HPC installations, offering them a
familiar interface to seamlessly exploit the raw computing power available.

HPK runs as a user-triggered service, instantiated via Slurm. Container workloads are handled by the hpk-kubelet
executable---a virtual Kubernetes node representing the entire cluster as a single entity. The
hpk-kubelet translates container lifecycle actions to Slurm scripts using Singularity/Apptainer commands.
HPK also includes several other customized Kubernetes modules to facilitate
integration with the HPC environment and simplify adoption by HPC centers.

\begin{acks}
The authors thankfully acknowledge the support of the European Commission under the Horizon Programme through project RISER (GA-101092993), as well as the European Commission and the Greek General Secretariat for Research and Innovation under the EuroHPC Programme through projects EUPEX (GA-101033975) and DEEP-SEA (GA-955606).
National contributions from the involved state members (including the Greek General Secretariat for Research and Innovation) match the EuroHPC funding.
\end{acks}

\bibliographystyle{ACM-Reference-Format}
\bibliography{main}

\end{document}